\documentclass[12pt]{article}
\usepackage{hyperref}
\usepackage{epsfig}
\usepackage{amsfonts} 
\usepackage{amsmath} 
\usepackage{amssymb}
\usepackage{color}

\def\half{{1 \over 2}}

\def\e{\eta}

\def\Or[#1]{{\text{O}}\left({#1}\right)}
\def\dotl[#1,#2]{\left\langle #1, #2 \right\rangle}
\def\dotlb[#1,#2]{[ #1, #2 ]}
\def\dotp[#1,#2]{(#1) \cdot (#2)}
\def\aff[#1,#2]{\hat{#1}(#2)}
\def\n4sym{{\cal N}=4 SYM}
\def\>{\rangle}
\def\<{\langle}

\def \half{{\textstyle {1 \over 2}}}
\def\weight[#1,#2,#3]{\{(#1),#2,#3\}}
\def\ads[#1]{$\text{AdS}_{#1}$}

\newcommand{\ba}{\begin{eqnarray}}
\newcommand{\ea}{\end{eqnarray}}

\newcommand{\nn}{\nonumber}
\newcommand{\beq}{\begin{equation}}
\newcommand{\eeq}{\end{equation}}

\newcommand{\be}{\begin{equation}}
\newcommand{\ee}{\end{equation}}  
\newcommand{\bi}{\begin{itemize}}
\newcommand{\ei}{\end{itemize}} 
\newcommand{\Tr}{{\rm Tr}}

\renewcommand{\bar}{\overline}

\renewcommand{\hat}{\widehat}

  \let\g=\gamma \let\d=\delta \let\e=\epsilon
    
  \let\n=\nu  
     
  \let\D=\Delta  
    
    \let\G=\Gamma

\renewcommand{\ba}{\begin{eqnarray}}
\renewcommand{\ea}{\end{eqnarray}}
\newcommand{\bea}{\begin{eqnarray}}
\newcommand{\eea}{\end{eqnarray}}

\begin{document}

\begin{center}
$$$$
{\Large\textbf{\mathversion{bold}
Skeleton expansion and large spin bootstrap
\\for $\phi^3$ theory 
}\par}

\vspace{1.0cm}

\textrm{Vasco Goncalves}
\\ \vspace{1.2cm}
\footnotesize{\textit{
ICTP South American Institute for Fundamental Research Instituto de Fisica Teorica,
UNESP - Univ. Estadual Paulista Rua Dr. Bento T. Ferraz 271, 01140-070, Sao Paulo,
SP, Brasil\\
}  }

\vspace{1.2cm}

\vspace{4mm}

\par\vspace{1.5cm}

\textbf{Abstract:}\vspace{2mm}
We study skeleton expansion of $\phi^3$ theory in $6+\epsilon$ dimensions as well as its global symmetry generalizations. We use it to compute the four point function of the scalar field $\phi$ up to $\epsilon^2$.
 
We also do a large spin bootstrap approach to the same model up to order $\epsilon^2$\\ and check that both results agree.

\end{center}

\noindent

\setcounter{page}{1}
\renewcommand{\thefootnote}{\arabic{footnote}}
\setcounter{footnote}{0}

\setcounter{tocdepth}{2}

 \def\nref#1{{(\ref{#1})}}

\newpage
\newpage

\tableofcontents

\parskip 5pt plus 1pt   \jot = 1.5ex

%%%%%%%%%%%%%%%%%%%%%%%%%%%%%%%%%%%%%%%%%%%%%%%%%%%%%%%%%%%%%%%
%\vfill\eject
\newpage
%%%%%%%%%%%%%%%%%%%%%%%%%%%%%%%%%%%%%%%%%%%%%%%%%%
\section*{Introduction}
%%%%%%%%%%%%%%%%%%%%%%%%%%%%%%%%%%%%%%%%%%%%%%%%%%

Over the last few years there has been a huge progress in the understanding of the properties of conformal field theories (CFTs). The use of unitarity, crossing symmetry and analyticity has allowed to derive many results that generic CFTs have to satisfy. Some examples of these properties are the presence of "double twist" operators \cite{Fitzpatrick:2012yx,Komargodski:2012ek}, the behavior of operators at large spin\cite{Alday:2013cwa}, Regge behavior\cite{Cornalba:2007fs,Caron-Huot:2017vep,Costa:2012cb,Kravchuk:2018htv} or the allowed singularities in a correlation function\cite{Maldacena:2015iua}. 

Crossing symmetry has also been explored in the analytical bootstrap of four point functions in an $\epsilon$ expansion. There are mainly two main methods, one o them tries to solve the bootstrap equations as a perturbation from large spin  operators\cite{Alday:2016njk} while in the other one tries to decompose a four point function in terms of exchange Witten diagrams\cite{Gopakumar:2016wkt,Gopakumar:2016cpb,Dey:2016mcs,Dey:2017oim,Dey:2017fab}. These two methods have the advantage that they only require consistency conditions (crossing, OPE decomposition,...) of the four point function. One of the most remarkable results is that it is possible to extract some OPE data from these conditions, for example it was possible to obtain the anomalous dimension of leading twist operators up to order $\epsilon^4$ \cite{Henriksson:2018myn}. However it is not straightforward to obtain arbitrary high orders in the coupling or to obtain OPE data of subleading twist operators because one has usually to deal with degenerate operators. 

On the other side there is the usual approach to correlation function where one needs to obtain all the diagrams, compute $\beta$ function, renormalize the operators and compute the integrals. This traditional approach is already a tour de force at low loop orders. In this paper we compute a four point function of a scalar field $\phi$ in $d=6+\epsilon$ up to order $\epsilon^2$ using a improved perturbation theory, called skeleton diagram expansion. In practice this assumes one is already at the fixed point of the theory and we only use in our computation physical quantities. In particular, one does not have to compute the $\beta$ neither we have to renormalize the operators. As will be shown in the main text the the number of diagrams to be computed are significantly less than the standard approach. Of course the method still requires the computation of Feynman integrals but fortunately there was a huge progress over the last few years in their evaluation. For example all that have been computed for this paper can be done using a computer package in less $2$ minutes on a normal laptop.  

The $\phi^3$ model has also been studied in the context of large spin perturbation in \cite{Alday:2015ewa} up to first order in $\epsilon$. We extend this result to order $\epsilon^2$ using large spin bootstrap and we find a perfect match with the one obtained from the skeleton diagram expansion. 

The basic idea of the skeleton diagram expansion  is explained in the next section. In section $3$ we compute the OPE decomposition of the four point function. In section $4$ we do the large spin pertubation for the $\phi^3$ model. In the last section we point out future direction and open problems. Usually one of the main difficulties in perturbative computations is the evaluation of Feynman integrals. We explain in detail in the appendices how this may be achieved .

%%%%%%%%%%%%%%%%%%%%%%%%%%%%%%%%%%%%%%%%%%%%%%%%%%
\section{Skeleton expansion}
%%%%%%%%%%%%%%%%%%%%%%%%%%%%%%%%%%%%%%%%%%%%%%%%%%
The standard approach to do perturbative computations in a conformal field theory requires the enumeration/evaluation of all relevant Feynman diagrams, the computation of the $\beta$ function(to flow to the fixed point) and removal of  divergences by renormalizing the bare operators\footnote{Let us remark that this is a necessary step even if the theory is already at the fixed point.}. 

The approach followed in this paper, skeleton diagram expansion\cite{Mack:1973kaa,Petkou:1994ad,Petkou:1995vu,Petkou:1996np}, bypasses the last two steps and reduces the number of diagrams that need to be computed. In this expansion we will use the exact two point function and  
 vertex of the theory.  In a standard perturbative expansion one has to consider diagrams that correct the propagator and cubic couplings. By using the exact propagator and vertex of the theory we assume that these diagrams were already package in these building blocks.

 Recall that in a conformal field theory, two and three point functions are given just in terms of two set of numbers, the dimension of the operator and the OPE coefficient
\begin{align}
\langle \phi_i(x)\phi_j(y)\rangle &=
\frac{\delta_{ij}}{(x-y)^{2\Delta_i}}\ ,\\
\langle \phi_1(x_1)\phi_2(x_2)\phi_3(x_3)\rangle &=
\frac{C_{123}}{x_{12}^{\Delta_{12,3}}x_{13}^{\Delta_{13,2}}x_{23}^{\Delta_{23,1}}},\, \ \ \ \ \Delta_{ij,k}=\Delta_i+\Delta_j-\Delta_k .
\end{align}
The exact three point cubic vertex can be extracted from the amputated three point function by
\be
\langle \phi_1(x_1)\phi_2(x_2)\phi_3(x_3)\rangle =
\int \bigg(\prod_{i=1}^3dy_i \langle \phi_i(x_i)\phi_i(y_i)\rangle \bigg)\,
V(y_1,y_2,y_3)\label{eq:CubicDressed}\ .
\ee
where the cubic vertex can be computed just by imposing that the integral is conformal
\be
V(y_1,y_2,y_3)=\frac{g_{123}}{(y_{12}^{2})^{\frac{d-\Delta_{12,3}}{2}}\,(y_{13}^{2})^{\frac{d-\Delta_{13,2}}{2}}\,(y_{23}^{2})^{\frac{d-\Delta_{23,1}}{2}}}
\label{dressedvertex}
\ee
where $g$ is a constant that is related with the OPE coefficient. The integrations in (\ref{dressedvertex}) can be computed using the star-triangle formula 
\begin{align}
\label{eq:StarTriangle}
\int \frac{d^dx_4}{
x_{14}^{2\Delta_1}x_{24}^{2\Delta_2}
x_{34}^{2\Delta_3}}=
\frac{\kappa(\D_1,\D_2,\D_3)}{
x_{12}^{\Delta_{12,3}}
x_{13}^{\D_{13,2}}
x_{23}^{\D_{23,1}}}\ ,
\end{align}
where $\sum \D_i =d$ and 
\be
\kappa(\D_1,\D_2,\D_3)=\pi^{\frac{d}{2}}\frac{
\G\left(\frac{\D_1+\D_2-\D_3}{2}\right)
\G\left(\frac{\D_1+\D_3-\D_2}{2}\right)
\G\left(\frac{\D_2+\D_3-\D_1}{2}\right)
}{
\G(\D_1)\G(\D_2)\G(\D_3)
}\ .
\ee
The relation between the OPE coefficient and the constant $g$ can be easily computed using the star-triangle formula
\begin{align}
\label{Ctog}
{\textstyle{C_{123}= g_{123}\, \kappa \big(\Delta _1,\frac{d -\Delta _{12,3}}{2},\frac{d -\Delta _{13,2}}{2}\big) \kappa \big(\frac{d -\Delta _{23,1}}{2},
\frac{d -\Delta _{13,2}}{2}
% \frac{d -\Delta _{23,3}}{2}
,\Delta _3\big) \kappa \big(
\frac{\Delta _{13,2}}{2},\Delta _2,\frac{2d-\Delta _1-\Delta _2-\Delta _3}{2}\big).}} 
\end{align}

All diagrams and sub-diagrams  that correspond to propagator  or vertex corrections are not allowed since their effect is already included in the building blocks. Another property of the expansion is that all integrals that appear are conformal and so they do not have any IR divergences, moreover all the integrals appearing in this paper are also UV finite.  

The existence of a small parameter in the theory may be used to truncate the number of diagrams that contribute to each order in a series expansion of this small parameter. A simple example where this happens is when the cubic coupling is proportional to this small parameter. One example where this happens is in $\phi^3$ theory where the coupling $g$ or OPE coefficient $C_{\phi\phi\phi}$ go to zero as $\epsilon\rightarrow 0$. The precise relation follows from (\ref{Ctog})
\begin{align}
g_{\phi\phi\phi} = \frac{C_{\phi\phi\phi}(\gamma_{\phi}-1)}{\pi^9}+O(\epsilon^3)
\end{align}
where we used that the dimension $d=6+\epsilon$ and that the dimension of the operator $\phi$ is given in terms of its free dimension $2$ and a small correction $\Delta_{\phi}=2+\epsilon \gamma_{\phi}$. 

Another example is scalar $O(N)$ models in $d$ dimension  for large $N$, in this case one has that the cubic coupling is proportional to $\frac{1}{N}$.

We have only computed four point functions of $\phi$ operators but in principle it should be possible to compute higher point functions as well. 

\begin{figure}
\begin{centering}
\includegraphics[scale=.85]{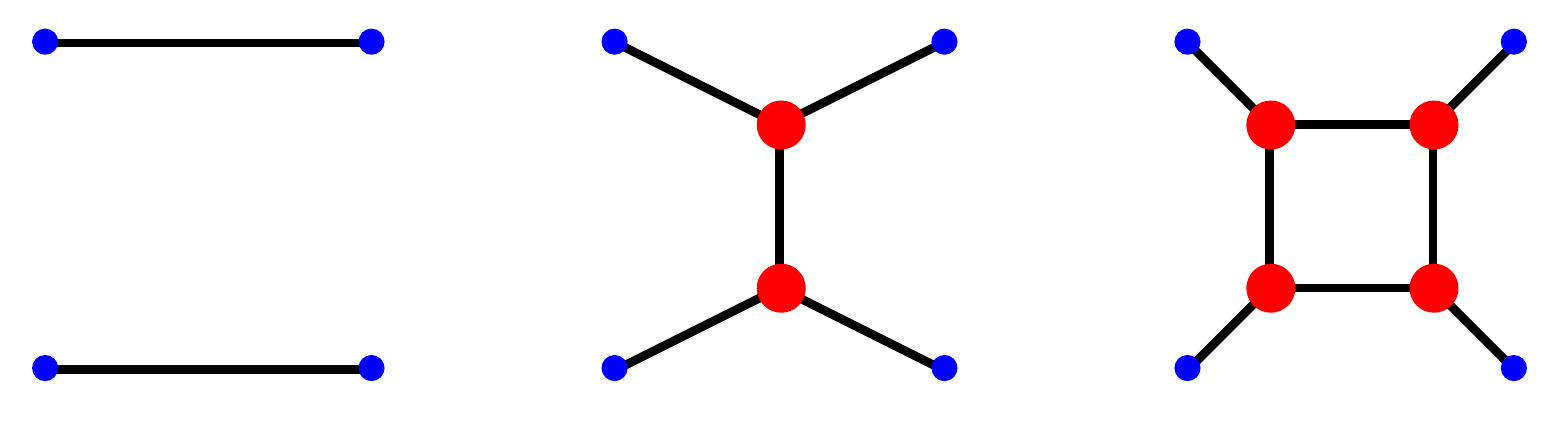}
\includegraphics[scale=.5]{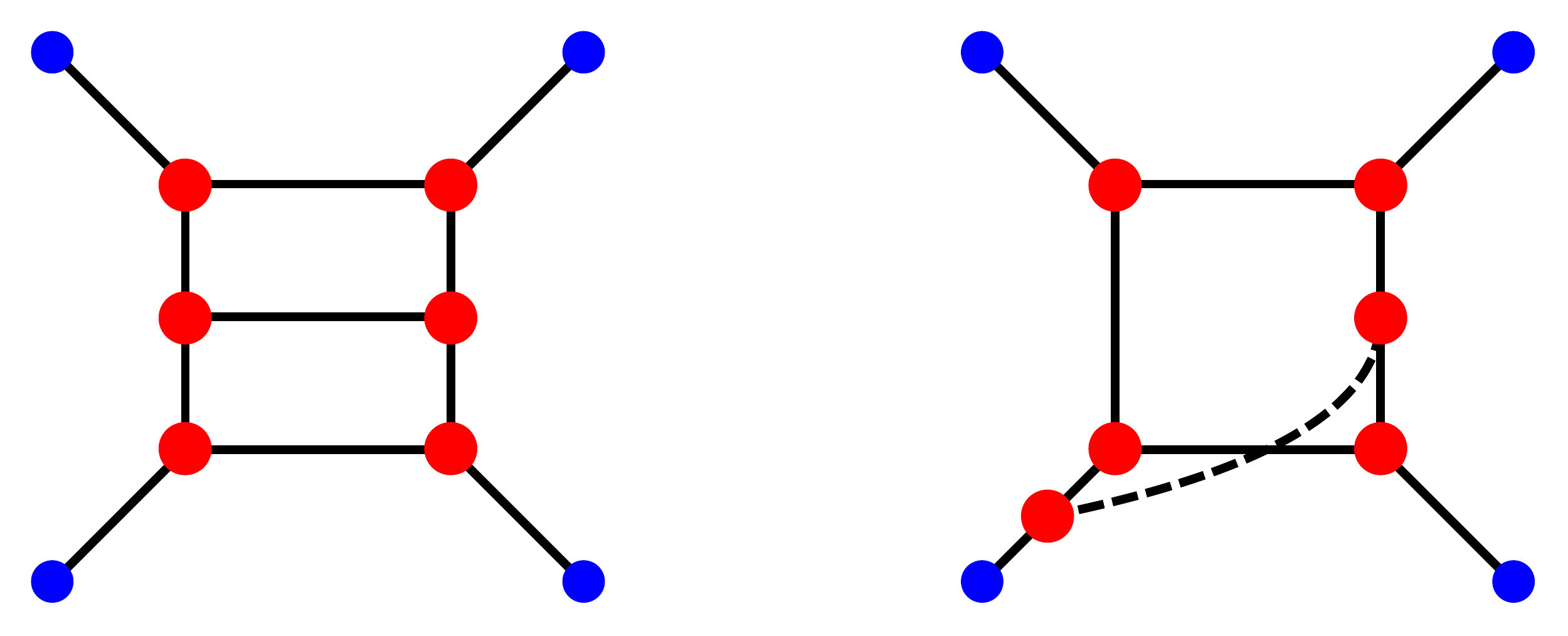}
\par\end{centering}
\caption{\label{fig:Skeleton4pt}
Skeleton graphs for four point function to order $g^6$. To these we must add the ones obtained by permutations of the external points, which are marked in blue. The red disks represent dressed vertices and the black lines represent full propagators. 
}
\end{figure}

The relevant diagrams for the four point function of $\phi$ up to order $\epsilon^3$ are shown in fig (\ref{fig:Skeleton4pt}) 
and it can be explicitly checked that there are no diagrams that correct the two point function or the three point vertex. 

\begin{figure}
\begin{centering}
\includegraphics[scale=.6]{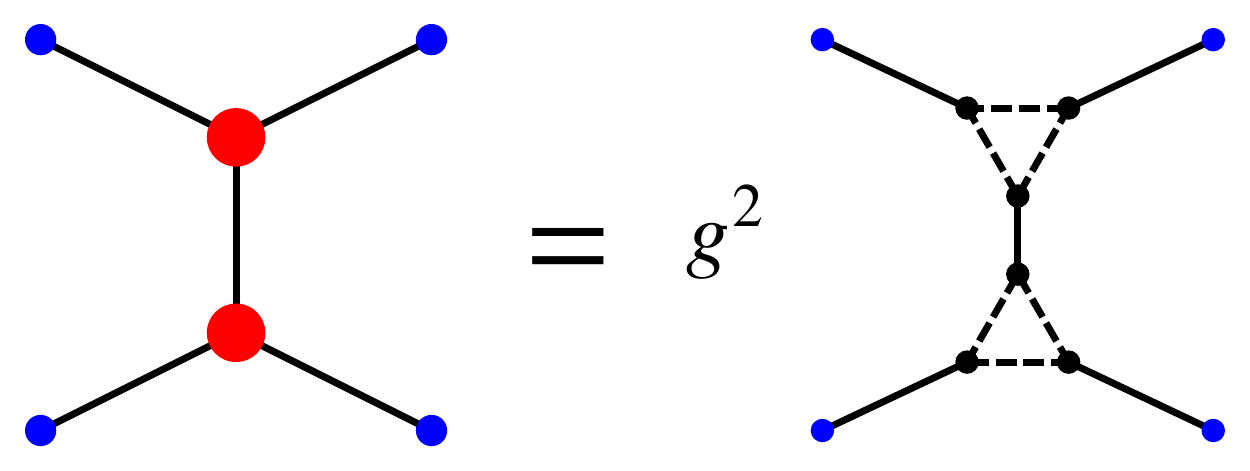}
\includegraphics[scale=.6]{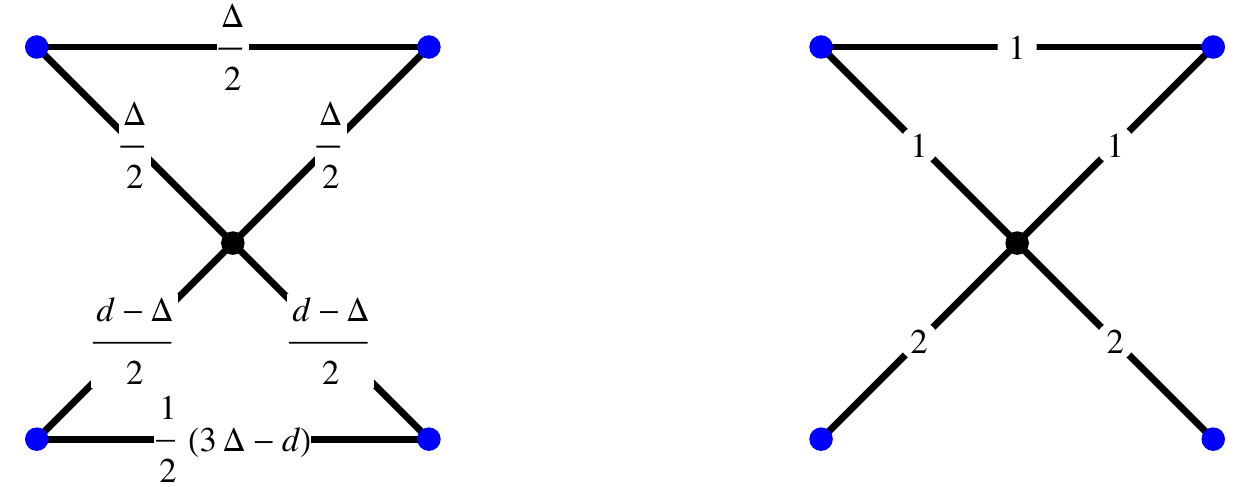}
\par\end{centering}
\caption{\label{fig:SimpSkeleton1}
Left: For each red dot should be replaced exact cubic vertex and integrated over three points. 
Right: Graph obtained by successive application of the star-triangle formula.
The dimension associated to each propagator is shown explicitly. The last image corresponds to the case with $\Delta=2$ and $d=6$}
\end{figure}

Each red shown in the diagram should be interpreted as a vertex with three where one integrates over the three point of the vertices. An explicit example of this is shown in figure (\ref{fig:SimpSkeleton1}) or in the following expression
\begin{align}
\textrm{second diagram of }(1) = g_{\phi\phi\phi}^2 \int \frac{d^dx_5\,\dots\,d^dx_{10}}{(x_{15}^2x_{27}^2x_{36}^2x_{49}^2)^{\frac{\Delta}{2}}(x_{57}^2x_{58}^2x_{69}^2x_{6\,10}^2x_{78}^2x_{8\,10}^2x_{9\,10}^2)^{\frac{d-\Delta}{2}}}.
\end{align}
All but one integration can be done using star-triangle formula (\ref{eq:StarTriangle}) leading to the one loop integral 
\begin{align}
&\textrm{second diagram}=\nonumber\\
&\frac{C_{\phi\phi\phi}^2  \Gamma (\Delta )
   \Gamma^2 \left(\frac{d-\Delta }{2}\right)}{ \pi ^{\frac{d}{2}}\Gamma^2
   \left(\frac{\Delta }{2}\right) \Gamma
   \left(\frac{d-2\Delta}{2}\right)}
   \frac{1}{x_{12}^{3\D-d} x_{34}^{\D}}
   \int \frac{d^dx_5}{
   x_{15}^{d-\D} x_{25}^{d-\D}x_{35}^{\D}x_{45}^{\D} }=\frac{C_{\phi\phi\phi}^2\G(\D)
   u^{\frac{d-\D}{2}}\bar{D}_{\frac{d-\D}{2},\frac{d-\D}{2},\frac{\D}{2},\frac{\D}{2}}(u,v)}{\G\big(\frac{d-2\D}{2}\big)\G^4\big(\frac{\Delta}{2}\big)(x_{12}^2x_{34}^2)^{\Delta}}\ .
   \label{firstconnectedgraph}
\end{align}
where the $\bar{D}$-functions are defined by, 
\be
\int \frac{d^d y}{\prod_{i=1}^4
(x_i-y)^{2\D_i}} = 
\frac{\pi^{\frac{d}{2}}}{\prod_{i=1}^4 \G(\D_i)}
\frac{x_{14}^{2(h-\D_1-\D_4)}
x_{34}^{2(h-\D_3-\D_4)}}{
x_{13}^{2(h-\D_4)}x_{24}^{2\D_2}}
\bar{D}_{\D_1\,\D_2\,\D_3\,\D_4}(u,v)
\ee
with $h=\half\sum\D_i=\frac{d}{2}$ and where $u$ and $v$ are the usual cross ratios
\begin{align}
u=z\bar{z}=\frac{x_{12}^2x_{34}^2}{x_{13}^2x_{24}^2},\, \ \ \ \ v=(1-z)(1-\bar{z})=\frac{x_{14}^2x_{23}^2}{x_{13}^2x_{24}^2}.
\end{align}
This integral is known for any $\Delta$ and $d$ as series expansion in small $u$ and $(1-v)$\footnote{This could be used, in particular, to extract information from the leading order four point function of scalar $O(N)$ model for large $N$.} \cite{Dolan:2000ut}. Morever, it is possible to compute the integral in closed form when the dimension of the propagators in the the integral are perturbatively close to positive integer values and when the dimension $d$ is close to an even integer. We include in an auxialliary a script that computes this integral up to an high order in $\epsilon$. 

All other diagrams in figure (\ref{fig:Skeleton4pt}) can be simplified in the same way, {\em {i.e.}} write the red dots as a integral over the three point vertex and then use start-triangle formula to simplify the integrals. The third diagram in  figure (\ref{fig:Skeleton4pt}) simplifes to 
\begin{align}
&\textrm{third diagram}=\frac{h(z,\bar{z})}{(x_{12}^2x_{34}^2)^{2\Delta}}=\nonumber\\
&\frac{C_{\phi\phi\phi}^4   \Gamma^4 (\Delta )
   \Gamma^8 \!\left(\frac{d-\Delta }{2}\right)}{
   \pi ^{2 d}\Gamma^8
   \!\left(\frac{\Delta }{2}\right) \Gamma^4
   \!\left(\frac{d-2\Delta}{2} \right)}
   \!\int\! \frac{d^dx_5d^dx_6d^dx_7d^dx_8}{
    \left(x_{15} x_{26}x_{37}x_{48}\right)^{3\D-d}\left(x_{18} x_{25}x_{36}x_{47}
    x_{56} x_{67}x_{78}x_{58}\right)^{d-\D} }.\label{eq:Twoloopintegralfull}
\end{align}
Notice that this diagram is proportional to $C_{\phi\phi\phi}^4$, so in $\phi^3$ theory it should kick in only at order $\epsilon^2$. Let us remark that we are not aware if there exists a known series expansion in $u$ and $(1-v)$ for this integral for generic value of $\Delta$ and dimension $d$. However, at leading order in $\epsilon$, the integrand simplifies considerably as the exponents of some propagators go to zero and thus one can use star-triangle a couple more times leading to   
\begin{align}
\frac{h(z,\bar{z})}{(x_{12}^2x_{34}^2)^2}= C_{\phi\phi\phi}^4\int \frac{d^6x_5d^6x_7}{\pi^6}\frac{1}{
   x_{15}^2 x_{35}^{2}x_{17}^2  x_{37}^{2} (x_{57}^{2}x_{25}^{2}x_{47}^{2})^2  }.\label{eq:twoloopintegralsimplified}
\end{align}

It is possible to derive a differential equation that this integral has to satisfy, notice that the integral simplifies after   applying $\nabla^2_{2}$ to the integrand since 
\begin{align}
\nabla^2_{2}\frac{1}{(x_{25}^2)^2}\propto \delta^{(6)}(x_{56})
\end{align}
which trivializes one of the integrals. We show in the appendix how this can be used to obtain the integral. 

We also provide a script  to evaluate the integral (\ref{eq:Twoloopintegralfull}) for higher order of $\epsilon$ using the package HyperInt\cite{Panzer:2015ida}. 

The last two diagrams in figure (\ref{fig:Skeleton4pt}) can be simplified in the same way. We show here the final result for assuming $\Delta=2$ and $d=6$
\begin{align}
&\frac{f(z,\bar{z})}{(x_{12}^2x_{34}^2)^2}=\int  \frac{d^6x_5d^6x_6d^6x_7}{( x_{26}^2x_{17}^2x_{56}^2x_{57}^2)^2x_{35}^2 x_{37}^2x_{45}^2x_{46}^2 x_{67}^2}\\
&\frac{g(z,\bar{z})}{(x_{12}^2x_{34}^2)^2}=\int\frac{d^6x_5d^6x_6d^6x_7}{x_{16}^2x_{17}^2x_{26}^2x_{27}^2x_{35}^2x_{36}^2x_{56}^2(x_{45}^2x_{57}^2x_{67}^2)^2} .\label{eq:3loopintegrals}
\end{align}
Notice that these two integrals also satisfy differential equations coming from applying $\nabla^2$ to the integrands. However we were not able to solve them. Nonetheless we have computed them using the package HyperInt. Once we got the solution we verified that the result for these two integrals satisfied the differential equations. 

The integrals appearing in this work are naturally written in terms  of hyperlogarithms\footnote{These are also called in the litterature as Goncharov logarithms} defined  as \cite{Panzer:2015ida}
\begin{align}
G_{\omega_{\sigma_1}\dots\omega_{\sigma_r}}(z)=\int_{0}^{z}\frac{dz_1}{z_1-\sigma_{1}}\int_{0}^{z_1}\frac{dz_2}{z_2-\sigma_{2}}\int\dots \int_{0}^{z_{r-1}}\frac{dz_r}{z_r-\sigma_{r}}.
\end{align}
One advantage of expressing the result in terms of these functions is that it is possible to do series expansions efficiently and it is simpler to consider analytic continuations. It is possible to do series expansions of hyperlogarithms using the package HyperInt. 

Putting everything together we arrive at the four point function of the field $\phi$ given by 
\begin{align}
\langle \phi\dots\phi \rangle &= 1+(z\bar{z})^{\Delta}+\left(\frac{z\bar{z}}{(1-z)(1-\bar{z})}\right)^{\Delta}\nonumber\\
&+\frac{C_{\phi\phi\phi}^2\Gamma(\D)}{\Gamma^4\left(\frac{\D}{2}\right)\Gamma\left(\frac{d-2\D}{2}\right)}\big[
(z\bar{z})^{\frac{d-\Delta}{2}} \, \bar{D}(z,\bar{z}) +\textrm{permutations}
\big]\nonumber\\
&+C_{\phi\phi\phi}^4\big(h(z,\bar{z})+\textrm{permutations}\big)
\end{align}
where 
\begin{align}
&u^{\frac{d-\Delta}{2}}\bar{D}(z,\bar{z})=-\frac{(2-z-\bar{z}) \phi^{(1)}(z,\bar{z})}{(z-\bar{z})^3}-\frac{\frac{\ln v (z+\bar{z}-2 u)}{u}+2 \ln u}{(z-\bar{z})^2}+O(\epsilon)\\
&h(z,\bar{z})=\frac{(z+\bar{z}-2 z \bar{z}) \phi^{(2)}(z,\bar{z})}{(z-\bar{z})^3}+\frac{(z-1) z \partial_{z}\phi^{(2)}(z,\bar{z})-(\bar{z}-1) \bar{z} \partial_{\bar{z}}\phi^{(2)}(z,\bar{z})}{(z-\bar{z})^2}
\end{align}
In the appendix we show how to generalize the construction of the four point function for the with a global symmetry.
%%%%%%%%%%%%%%%%%%%%%%%%%%%%%%%%%%%%%%%%%%%%%%%%%%
\section{Conformal block analysis}
%%%%%%%%%%%%%%%%%%%%%%%%%%%%%%%%%%%%%%%%%%%%%%%%%%
One of the main results of the last section was the the construction of four point functions using skeleton digrams. The correlators are expressed in terms of some yet to be determined constants, namely, the dimensions of some operators and their OPE coefficients. These may be obtained by imposing that the four point function has a consistent OPE decomposition. 

The simplest example where a consistent OPE decomposition is able to fully determine all the undetermined constants is in $\phi^3$ in $d=6+\epsilon$ dimensions. The four point function of four $\phi$ is a non-trival function of the cross ratios $u$ and $v$
\be
\langle \phi(x_1)\phi(x_2)\phi(x_3)\phi(x_4)\rangle = \frac{\mathcal{A}(u,v)}{x_{12}^{2\Delta}x_{34}^{2\Delta}}
\ee
The OPE between the operators at positions $x_1$ and $x_2$ leads to the following conformal block decomposition for this correlator
\begin{align}
g(u,v) = \sum_{\mathcal{O}}c_{\phi\phi \mathcal{O}}^2G_{\Delta_{\mathcal{O}},l}(u,v)
\end{align}
where $c_{\phi\phi \mathcal{O}}$, $\Delta_{\mathcal{O}}$ and $l$ are the OPE coefficient, dimension and spin of the operator $\mathcal{O}$. The conformal blocks $G_{\Delta,J}(u,v)$ are not known in closed form for generic dimension $d$. However we will use a series expansion in small  $u$ and fixed $v$ of the conformal blocks
\be
G_{\D,l}^d(u,v)=\sum_{m=0}^{\infty}u^{\frac{\D-l+2m}{2}}g_m(v),\, g_0(v)=\left(\frac{v-1}{2}\right)^l\!_2F_1\left(\frac{\Delta+l}{2},\frac{\Delta+l}{2},\Delta+l\right) \label{eq:ConfReggeBlocks}
\ee
where the functions $g_m(v)$ satisfy a differential recursion relation as shown in \cite{Costa:2012cb}. The precise form  of these functions is worked out in the appendix. 

It is instructive to analyze in more detail the first two orders in $\epsilon$ to understand how the undetermined coefficients may be fixed. The tree level four point function is given by 
\be
\mathcal{A}(z,\bar{z})=1+(z\bar{z})^2 + \left(\frac{z\bar{z}}{(1-z)(1-\bar{z})}\right)^2+O(\epsilon).
\ee
Notice that the minimal twist ($\textrm{twist}=\tau=\Delta-l$) that can flow at the order is $4$ as one can check by the leading power of $\bar{z}$ in the limit $\bar{z}\rightarrow 0$. This can also be shown by doing conformal block decomposition,
\be
\mathcal{A}(z,\bar{z})=G_{0,0}^6(z,\bar{z})+ \sum_{l =0 \atop {\rm even}}^\infty p_{0,l}\, G_{4+l,l}^6(z,\bar{z})
\label{zeroCBEwrong}
\ee
where
\be
p_{0,l}= \frac{2^l(l+2)!(l+1)!}{(2l+1)!}\ .
\ee
and the term $G_{0,0}^6$ is the contribution of the identity operator. 

At next order in $\epsilon$ the $\bar{D}$ functions start to contribute and one of them has the following behavior in the OPE limit
\begin{align}
\frac{\Gamma(\D)u^{\frac{d-\Delta}{2}}}{\Gamma^4\left(\frac{\D}{2}\right)\Gamma\left(\frac{d-2\D}{2}\right)}\bar{D}(u,v)\approx u+O(u)+O(1-v)
\end{align}
which is  consistent with the presence of an scalar operator with dimension $2$ appearing in the OPE. This is no surprise since the operator $\phi$ should start contributing at order $\epsilon$. What is more is that after doing the conformal block decomposition the scalar operator with dimension $4$ disappears from the spectrum. The reason for this is that an operator which was a primary at tree level became a descendent at order $\epsilon$, which is consistent if one thinks of the equations of motion for the field $\phi$\cite{Anselmi:1998ms,Henn:2005mw,Rychkov:2015naa}. At the level of the conformal blocks what is happening is the following 
\be
 G_{4,0}^6(u,v) 
= 6 \g_1 
\lim_{\e \to 0} \e\, G_{\D,0}^{6+\e}(u,v) \label{eq:FromPrimaryToDescent}
\ee
which automatically fixes the value of the undetermined constants. This pattern repeats itself at higher orders in $\epsilon$, thus determining all the values that were unfixed. 

Having the four point function at order $\epsilon^2$ gives us information about the anomalous dimension and OPE coefficients of leading and subleading twist operators. For the scalar $\phi$ we obtain
\begin{align}
&\D= \frac{d-2}{2} +\frac{\e}{18}-\frac{43}{1458} \e^2+\epsilon^3(\frac{8375}{472392}+O(\e^3),\, \\
& C_{\phi\phi\phi}^2=  \frac{2}{3}\e
 -\frac{143}{243}\e^2 +O(\e^3).
\end{align}
which agrees with the results previously computed\cite{deAlcantaraBonfim:1980pe,Gracey:2015tta}. 
For example, the twist four operator come from start contributing at order $u^2$. The corresponding anomalous dimension and OPE coefficient can be extracted at order $\epsilon$ from the $u^2$ term in the four point function
\begin{align}
&\D_l= 2\D+l  +\gamma_l^{(1)}\e+\gamma^{(2)}_l\e^2, \ \  \,\gamma_l^{(1)}=-\frac{4}{3(l+1)(l+2)}\label{eq:DimensionsofOperators} \\
&\e^2\gamma_l^{(2)} = \frac{2 \left(86-l (l (l (127 l+492)+382)-177)\right)}{243 (l+1)^3 (l+2)^3}-\frac{5\gamma_l^{(1)}S_l}{9}\nonumber
\end{align}
and 
\begin{align}
p_{0,l}= \frac{2^{l+1}(\Delta)_l^2}{l!(\Delta_l-1)_l}(1+\e\, \mathcal{P}_l^{(1)})+O(\e^2),\,\mathcal{P}_l^{(1)} = -\frac{\gamma_l^{(1)}}{l+2}.
\end{align}
We did not try hard enough to find an analytic expression for generic spin for the higher order corrections to the twist $4$ operators. We present in the following table the values for the first few spins
\beq
\!\!\!\!\!\!\!\! \begin{array}{c|l}
l &  \ \ \  \text{Anomalous dimension for twist $4$} \vspace{0.1cm}
%\color{white}{\Big(_{2_{2_2}}}
\\ \hline
% 0 & 1 \color{white}{\Big(}  \\
2 &-\frac{\epsilon }{9}+\frac{43 \epsilon ^2}{729}+\dots \color{white}{\Big(} \\
4 &-\frac{2 \epsilon }{45}+\frac{2762 \epsilon ^2}{91125}+\dots  \color{white}{\Big(}\\
6 &-\frac{\epsilon }{42}+\frac{102005 \epsilon ^2}{5334336}+\dots  \color{white}{\Big(}\\
8 &-\frac{2 \epsilon }{135}+\frac{307553 \epsilon ^2}{22963500}+\dots  \color{white}{\Big(}\\
10 &-\frac{\epsilon }{99}+ \frac{181141 \epsilon ^2}{18112248}\dots \color{white}{\Big(}
\end{array}  \nn
\eeq

\beq
\!\!\!\!\!\!\!\! \begin{array}{c|l}
l &  \ \ \  \text{OPE coefficient of twist four operators for  $l$} \vspace{0.1cm}
%\color{white}{\Big(_{2_{2_2}}}
\\ \hline
% 0 & 1 \color{white}{\Big(}  \\
2 &\frac{24}{5}+\frac{634 \epsilon }{225}-\frac{26591 \epsilon ^2}{364500}+\dots \color{white}{\Big(} \\
4 &\frac{80}{21}+\frac{69736 \epsilon }{19845}+\frac{246255764 \epsilon ^2}{281302875}+\dots  \color{white}{\Big(}\\
6 &\frac{896}{429}+\frac{433174024 \epsilon }{173918745}+\frac{2047344914778847 \epsilon ^2}{1974210806868300}+\dots  \color{white}{\Big(}\\
8 &\frac{2304}{2431}+\frac{7479023872 \epsilon }{5584724145}+\frac{699105791225392786 \epsilon ^2}{962231914101583125}+\dots  \color{white}{\Big(}\\
\end{array}  \nn
\eeq
This family of primaries do not exhaust the operators that contribute to the four point function. For instance, there is a family of twist $6$ operators that start contributing at $u^3$. Their OPE coefficient starts at order $\epsilon$ and reads\footnote{Let us remark that twist $6$ are degenerate, so this OPE coefficient  should be interpreted as a sum rule for the operators that have the same spin and twist. }
\begin{align}
p_{1,l}= \epsilon\frac{(1+l)(4+l) (18+l (5+l)) (1+l)!^2}{216 (3+l)(3+2 l)!}.
\end{align}

The analysis at next leading order in $\epsilon$ is similar. There are corrections to the anomalous dimension and OPE coefficients of  the elementary field $\phi$, the twist four and six operators. At this order one starts also to see the contribution of operators with twist eight. 
\beq
\!\!\!\!\!\!\!\! \begin{array}{c|l}
l &  \ \ \  \text{Average of anomalous dimension for twist $6$ } \vspace{0.1cm}
%\color{white}{\Big(_{2_{2_2}}}
\\ \hline
% 0 & 1 \color{white}{\Big(}  \\
0 &\ \ \ \  O(\epsilon^2) +\dots \color{white}{\Big(} \\
2 &\ \ \ \ \frac{28}{45}\epsilon+\dots  \color{white}{\Big(}\\
4 &\ \ \ \ \frac{262}{315}\epsilon+\dots  \color{white}{\Big(}\\
6 & \ \ \ \ \frac{12353}{13230}\epsilon+\dots  \color{white}{\Big(}\\
8 &\frac{89648}{90585}\epsilon\dots \color{white}{\Big(}
\end{array}  \nn
\eeq

%%%%%%%%%%%%%%%%%%%%%%%%%%%%%%%%%%%%%%%%%%%%%%%%%%%%%%%%%%%%%%%%%%%%%%%%%%%%%%%%%%%%%%%%%%%%%%%%%%%%%%%%%%%%%%%%%%%%%
\subsection*{Leading Regge trajectory and inversion formula}
%%%%%%%%%%%%%%%%%%%%%%%%%%%%%%%%%%%%%%%%%%%%%%%%%%%%%%%%%%%%%%%%%%%%%%%%%%%%%%%%%%%%%%%%%%%%%%%%%%%%%%%%%%%%%%%%%%%%%
The leading Regge trajectory is defined as the family of operators that have the lowest twist per spin. In the present case these operators have twist $4$ , are non degenerate and their anomalous dimension is given by (\ref{eq:DimensionsofOperators}). The spin $2$ operator of this family is the stress energy tensor and one can verify that $\Delta_{2}=d$ which should happen since the dimension of the stress tensor is fixed. There is no spin $0$ operator in this family however our expression is analytic in the spin and one may compute the analytic continuation of the anomalous dimension to spin $0$. Curiously the analytic continuation to $0$ satisfies 
\begin{align}
\Delta_{0}-(d-\Delta_{\phi})=0, 
\end{align}
up to the order we have analytic data. This implies that the analytic continuation of the leading Regge trajectory down to spin $0$ is the shadow of the field $\phi$.

%%%%%%%%%%%%%%%%%%%%%%%%%%%%%%%%%%%%%%%%%%%%%%%%%%%%%%%%%%%%%%%%%%%%%%%%%%%%%%%%%
%\subsection*{Comment on bootstrapping the final result}
%%%%%%%%%%%%%%%%%%%%%%%%%%%%%%%%%%%%%%%%%%%%%%%%%%%%%%%%%%%%%%%%%%%%%%%%%%%%%%%%%
%We used the skeleton expansion to obtain the four point function up to order $\epsilon^2$. This approach  has lead to an expression that only depended on two unfixed coefficients $\Delta_{\phi}$ and  $C_{\phi\phi\phi}$. These were later determined by a OPE analysis. 
%
%One might wonder if the coefficients multiplying the integrals could be arbitrary. A simple analysis has shown us that the consistency of the four point function with previous orders in $\epsilon$ constraints the values of the coefficients that multiply the integrals. 
%
%All the integrals appearing in this work were finite and conformal. It would be tempting to consider a basis of finite conformal integrals and try to bootstrap the result using consistency conditions that a CFT has to satisfy\footnote{Let us remark that we are not stating that any perturbative CFT can be written in terms of finite conformal integrals. }. One could even use the analytic bootstrap results to fix completely the answer. 

%%%%%%%%%%%%%%%%%%%%%%%%%%%%%%%%%%%%%%%%%%%%%%%%%%%%%%%%%%%%%%%%%%%%%%%%%%%%%%%%%
\section{Large spin bootstrap}
%%%%%%%%%%%%%%%%%%%%%%%%%%%%%%%%%%%%%%%%%%%%%%%%%%%%%%%%%%%%%%%%%%%%%%%%%%%%%%%%%
In the previous section we have obtained the four point function of the field $\phi$ up to order $\epsilon^2$ using  an approach based on Feynman graph computation. Recently a  different approach based on consistency conditions coming from the bootstrap equations in the large spin limit was developed and applied to several models. In this section we do a large spin analysis to the $\phi^3$ model  similar to the one that was recently performed up to order $\epsilon^4$ in the $\phi^4$ \cite{Alday:2017vkk} with the some small differences. 

The four point function can be decomposed in terms of conformal blocks in either the $(12)$ channel or in the $(23)$ since the OPE is associative. The bootstrap equation is just the statement that both representations are equal
\begin{align}
\mathcal{A}(u,v)=\frac{u^{\Delta_\phi}}{v^{\Delta_\phi}}\mathcal{A}(v,u) \leftrightarrow \sum_{\Delta_k}C_{\phi\phi \mathcal{O}_k}^2G_{\Delta_k,l_k}(u,v)=\frac{u^{\Delta_\phi}}{v^{\Delta_\phi}}\sum_{\Delta_k}C_{\phi\phi \mathcal{O}_k}^2G_{\Delta_k,l_k}(v,u).\label{eq:bootstrapequations}
\end{align}
The simplicity of the equations contrasts with the restrictive space of solutions that exists. We will focus on theories that are perturbatively close to $d=6+\epsilon$ and whose fundamental field is also perturbatively close to the free one 
\begin{align}
\Delta_{\phi}=\frac{d-2}{2}+\epsilon\gamma_{\phi}^{(1)}+\epsilon^2\gamma_{\phi}^{(2)}+\dots.
\end{align}
and with OPE coefficient $C_{\phi\phi\phi}^2\sim \epsilon+O(\epsilon)$. Moreover, we will assume that the twist $6$  operators only couple to the $\phi$ field at order $\epsilon$. 

The methods of large spin perturbation theory were developed in \cite{Alday:2016njk}. The main idea is to derive constraints from (\ref{eq:bootstrapequations}) starting from the region $(u,v)=(0,0)$ in cross ratio space. The name large spin comes from the fact that the dominant contributions to this equation comes from operators with large spin as can be seen from the singularity structure of the conformal blocks
\begin{align}
G_{\Delta,l}(u,v)\approx  \frac{\Gamma\left(\Delta+l\right)}{\Gamma\left(\frac{\Delta+l}{2}\right)}u^{\frac{\Delta-l}{2}}\ln v +\dots
\end{align}
and bootstrap equations in this region\cite{Alday:2013cwa}. More concretely a single term in the conformal block decomposition of the lhs of (\ref{eq:bootstrapequations}) diverges at most with a $\ln v$ singularity while the rhs has a power law divergence. The apparent mismatch of the singularity structure of the equation is fixed after summing over an infinite family of operators with the same twist and different spin. Notice that this singularity enhancement needs to come from the tail of the sum since a finite sum of terms will never generate a power law divergence.

It is thus natural to write the anomalous dimension and OPE coefficient as an expansion around large spin. It turns out, as shown in \cite{Alday:2015ewa}, that this expansion can be written in terms of inverse powers of the conformal spin  $J^2=(2l+\tau)(2l+\tau-2)$ where $\tau$ is the twist. The twist conformal block, defined in 
 \cite{Alday:2016njk,Alday:2016jfr} are kinematical objects that make the analysis of the equations more systematic
\begin{align}
H^{(m)}_{\tau}(u,v)=\sum_{l}\frac{a_{\tau,l}^{(0)}}{J^{2m}}u^{\frac{\tau}{2}}g_{\tau,l}(u,v),\, \ \ \ \ a_{l}^{(0)}=\frac{2\Gamma^2(d/2+l-1)\Gamma(d+l-3)}{\Gamma^2(d/2-1)\Gamma(l+1)\Gamma(d+2l-3)}\label{eq:twistconformalblocks}
\end{align}
where the $a_{l}^{(0)}$ are just OPE coefficients. Twist conformal blocks satisfy a recursion relation in $m$ 
\begin{align}
&\mathcal{C}H^{(m+1)}_{\tau}(u,v)=H^{(m)}_{\tau}(u,v),\, \ \ \mathcal{C}=D+\bar{D}+\\
+\frac{(d-2)z\bar{z}}{z-\bar{z}}&((1-z)\partial_z-(1-\bar{z})\partial_{\bar{z}})+\frac{\tau(2d-\tau-2)}{4},\, \ \ \ \ \ D=(1-z)z^2\partial^2-z^2\partial\label{eq:differentialrecurrencerelation}.
\end{align}
since $u^{\frac{\tau}{2}}g_{\tau,l}(u,v)$ is an eigenfunction of $\mathcal{C}$ with eigenvalue $J^2$. On the left hand side of the bootstrap equation (\ref{eq:bootstrapequations}) there is a scalar operator with dimension $\Delta_{\phi}$ and then an infinite family of operators with twists $2\Delta_{\phi}+2m+\gamma_{l,m}$.

The limit $u\rightarrow 0$ singles the operators with lowest twist. We are now equipped  to analyze the bootstrap equations (\ref{eq:bootstrapequations})
\begin{align}
&C^2_{\phi\phi\phi}G_{\Delta_{\phi},l}(u,v)+\sum_{l}C_{\phi\phi\mathcal{O}_{2+l,l}}^2G_{\Delta,l}(u,v)\nonumber\\
&=\frac{u^2}{v^2}\left(1+\frac{\epsilon(1+2\gamma_{\phi}^{(1)})}{2}\ln \frac{u}{v}+\frac{\epsilon ^2}{8} \ln \frac{u}{v} \left(8 \gamma_{\phi}^{(2)}+(2 \gamma_{\phi}^{(1)}+1)^2 \ln \frac{u}{v}\right)+\right)\mathcal{A}(v,u)\label{eq:bootstrapequationexpanded}
\end{align}
where $\mathcal{A}(v,u)$ stands for the t-channel conformal block decomposition. Now one tries to decompose the left hand side in terms of twist conformal blocks. This problem has already been studied up to order $\epsilon$ for $\phi$ in \cite{Alday:2016jfr}.

At tree level only the seed twist conformal block $H_{d-2}^{0}$
\begin{align}
H_{d-2}^{0} = u^{\frac{d-2}{2}} + \left(\frac{u}{v}\right)^{\frac{d-2}{2}}
\end{align}
contribute and it is clear that this matches the known result for the four point function at $\epsilon^0$. 

At order $\epsilon$ the operator $\phi$ starts to contribute to the four point function and so its contribution must be included in the rhs of (\ref{eq:bootstrapequations}) as it will give another power law singularity
\begin{align}
&\sum_{l}C_{\phi\phi\mathcal{O}_{2+l,l}}^2G_{\Delta,l}(u,v)\nonumber\\
&=\frac{u^2}{v^2}\left(1+\frac{\epsilon(1+2\gamma_{\phi}^{(1)})}{2}\ln \frac{u}{v}\right)\left(1+C_{\phi\phi\phi}^2G_{\Delta_{\phi},0}(v,u)+\dots\right)\label{eq:bootstrapequation3}
\end{align}
where the $\dots$ represent both less divergent terms in $v$ and higher order in $\epsilon$ and where we neglected the contribution of the operator $\phi$ on the right hand side because it does give an enhanced singularity.\footnote{The reader might wonder why we do not include the contribution of twist $4$ operators to $\mathcal{A}(v,u)$. One reason is that they give a less divergent contribution. This can be easily from the inversion formula derived in \cite{Caron-Huot:2017vep}, in fact in \cite{Alday:2017vkk} it was shown that the large spin method is equivalent to the CFT inversion formula. The dynamical data is given by a double discontinuity around $\bar{z}=1$ which kills the $\ln v$ term that comes from the correction of twist $4$ operators in the $t$ channel block. Another reason is that they don't generate power law singularities in the rhs of (\ref{eq:bootstrapequations}) } The scalar conformal block that enters in the $t$ channel decomposition is known in terms of a double series or single sum over hypergeometric function
\begin{align}
G_{\Delta,0}(v,u)&=\sum_{m,n=0}\frac{(\Delta/2)_m^2(\Delta/2)^2_{m+n}}{m!n!(\Delta+1-d/2)_{m}(\Delta)_{2m+n}}v^m(1-u)^n,\,\\
&=\sum_{m=0}^{\infty}\frac{v^m \left(\frac{\Delta }{2}\right)_m^4 \, _2F_1\left(m+\frac{\Delta }{2},m+\frac{\Delta }{2},2 m+\Delta ,1-u\right)}{m! (\Delta )_{2 m} \left(\Delta +1-\frac{d}{2}\right)_m}
\end{align}
which is more suitable to analyze the expansion around $(u,v)\approx (0,0)$ and $\epsilon\approx 0$. First we use a well known formula to expand the hypergeometric function around $1$, then we series expand in $\epsilon$ around $0$. The sum over $m$ can then be recognized as an hypergeometric function
\begin{align}
&G_{\Delta,0}(v,u)\approx v^{\frac{\Delta}{2}}\bigg[-\frac{\ln u\Gamma (\Delta ) \, _2F_1\left(\frac{\Delta }{2},\frac{\Delta }{2},\Delta+1-\frac{d}{2},v\right)}{\Gamma \left(\frac{\Delta }{2}\right)^2}+\nonumber\\
&(\partial_{c}+\gamma_{e})\frac{2^{\Delta } \Gamma \left(\frac{c}{2}\right)^2 \Gamma \left(\frac{\Delta +1}{2}\right) \, _2F_1\left(\frac{c}{2},\frac{c}{2},\Delta+1-\frac{d}{2},v\right)}{\sqrt{\pi } \Gamma \left(\frac{\Delta }{2}\right)^3}\bigg|_{c=\Delta}+\dots\bigg]
\end{align}
where the dots represent higher order terms in $u$. So from here one sees that there is another a power law singularity in the right hand side of (\ref{eq:bootstrapequation3}) which is matched by the contribution of a twist conformal block $H_{d-2}^{1}$.  It is important to remark that one does not need to know the twist conformal blocks in detail since we are only interested in the limit $(u,v)\approx (0,0)$. The twist conformal block  $H_{d-2}^{1}$ can be obtained from the differential recurrence relation (\ref{eq:differentialrecurrencerelation}) as explained in \cite{Alday:2016jfr}
\begin{align}
H_{d-2}^{1}(u,v)\approx \frac{u^{\frac{d-2}{2}}\Gamma\left(\frac{d-4}{2}\right)}{(1-u)v^{\frac{d-4}{2}}\Gamma^2\left(\frac{d-2}{2}\right)}. 
\end{align}
The part containing $\frac{\epsilon(1+2\gamma_{\phi}^{(1)})}{2}\ln \frac{u}{v}$ can be recovered from the bootstrap equation (\ref{eq:bootstrapequation3}) just by a constant term in the anomalous dimension of twist $4$ operators. This leads to an anomalous dimension 
\begin{align}
\gamma_{l}=2\gamma_{\phi}-\frac{2C_{\phi\phi\phi}^2}{(l+2)(l+1)}.
\end{align}
Remember that both $C_{\phi\phi\phi}$  and $\gamma_{\phi}$ are unknowns(from the perspective of this bootstrap exercise)  up to this point but they can be 
fixed by requiring the conservation of the stress energy tensor, $l=2$, and that the analytic continuation of the anomalous dimension up to spin zero is the shadow of the operator $\phi$. 

Before moving on to order $\epsilon^2$ it is important to emphasize a few points. Both the anomalous dimension and the OPE coefficients of twist four operators were obtained from the bootstrap equation at order $\epsilon$ from the contribution of the operator $\phi$ in the $(23)$ channel of the conformal bock decomposition. This was a special feature that will cease at higher orders. For instance at order $\epsilon^2$ the $(23)$ conformal decomposition will have a divergence proportional to $\ln^2v$ coming from the twist $4$ operators which cannot be matched by a finite number of terms in the right hand side of the $(12)$ channel\footnote{Twist $6$ and $8$ operators are also present in the four point function but they only give at most $\ln v$ and $\ln^0 v$ divergences  so they do not have to be taken into account for the reasons mentioned above.}.  This will lead to more complex functions on the rhs of the bootstrap equation and consequently the decomposition in terms of twist conformal blocks would get more involved. Fortunately in \cite{Alday:2017vkk} it was shown how to obtain the OPE data in terms of an integral transform of the rhs of (\ref{eq:bootstrapequations}) 
\begin{align}
a(J)=\frac{2\bar{h}-1}{\pi^2}\int_{0}^{1}dt d\bar{z}\frac{\bar{z}^{\bar{h}-2}(t(1-t))^{\bar{h}-1}}{(1-t\bar{z})^{\bar{h}}}\textrm{dDisc}\mathcal{A}(\bar{z})\label{eq:doublediscontinuity}
\end{align}
where we used only the limit $z\rightarrow 0$ of the four point function\footnote{ This is the same expression as presented in \cite{Alday:2017vkk} which was used in $d=4-\epsilon$. The reader might wonder why we are allowed to use the same equation to our case in $d=6+\epsilon$ without any modification. The reason is that we are focusing on the leading twist family of operators the conformal blocks in the limit of $z\rightarrow 0$ do not depend on the dimension of the theory one is studying. } and where $\bar{h}$ is defined by $J^2=\bar{h}(\bar{h}-1)$. The function  $a(J)$ represents the product of the anomalous dimension to some power and the OPE coefficient that comes from the conformal block decomposition of the $(12)$ channel
\begin{align}
u^{2}\sum  \ln^n u C_{\phi\phi \mathcal{O}}(J) \gamma^n(J)g_0(\bar{z}).
\end{align}
The OPE data in (\ref{eq:doublediscontinuity}) is written in terms of a double discontinuity defined as
\begin{align}
\textrm{dDisc}\mathcal{A}(\bar{z})=\mathcal{A}(\bar{z}) - \frac{1}{2}\mathcal{A}^\circlearrowleft(\bar{z})-\frac{1}{2}\mathcal{A}^\circlearrowright(\bar{z})
\end{align}
where $\mathcal{A}^{\circlearrowleft,\circlearrowright}$  represents the analytic continuation around $\bar{z}=1$ clockwise and counterclockwise. One interesting property of the double discontinuity is that it kills  $\ln 1- \bar{z}$ as well as regular terms around $\bar{z}=1$. This means that one can reproduce the order $\epsilon^2$ of $a(J)$ using the one loop OPE data in the conformal block decomposition of the $(23)$ since higher order corrections would only contribute to the $\ln 1-\bar{z}$ and regular terms.   

We have have computed $\textrm{dDisc}\mathcal{A}(\bar{z})$ from the $(23)$ conformal block decomposition then we have used a table from \cite{Alday:2017vkk} to compute the integrals and obtain the function $a(J)$, which agreed with the OPE data from the skeleton diagram approach. 

%%%%%%%%%%%%%%%%%%%%%%%%%%%%%%%%%%%%%%%%%%%%%%%%%%%%%%%
\section{Conclusions and future directions}
%%%%%%%%%%%%%%%%%%%%%%%%%%%%%%%%%%%%%%%%%%%%%%%%%%%%%%%
We have computed the four point function of the operator $\phi$ in $\phi^3$ theory in $d=6+\epsilon$ dimensions up to order $\epsilon^2$. Moreover, the  approach followed here used only physical observables as the dimension and OPE coefficients of the operators. We also obtained anomalous dimensions and OPE coefficents of leading and subleading twist operators up to order $\epsilon^2$. It was also shown how to generalize these results to the case with a global symmetry group. 

Two obvious generalizations that were not attempted are the computation to higher order in $\epsilon$ or the construction of higher point. The integrals appearing in each of these objects should fall in the class of integrals where the package HyperInt can trivialize its evaluation. 

This method can also be applied to other theories, for example scalar large N $O(N)$ theories in $d$ dimensions or generalized free field theories. It would also be interesting to study skeleton expansion for $\phi^4$ in $d=4-2\epsilon$ dimensions or conformal QED\footnote{The skeleton expansion used only physical data  in the building blocks and also its conformal structure. In conformal QED is a gauge theory and so one has to figure out how to do the expansion in this case. For example, the propagator of the gauge field only has conformal structure for a specific gauge. }. Another approach to obtain these theories is to write an ansatz in terms of finite conformal integrals and try to bootstrap the result.  
One could also try to do skeleton expansions for defect conformal field theories. 

We did not explore Mellin amplitudes in the context of the skeleton expansion. It is known that this language simplifies the analysis the four point function in many some cases. One could try to compute what is the Mellin amplitude for this correlator and check if there is a simplification.

\section*{Acknowledgements} 
We would like to thank Joao Penedones for suggesting and collaborating in a "early" stage of the project (and also many useful discussions). We would like to thank Emilio Trevisani and Marco Meineri for collaborating in a "early" stage of the project. We are grateful to Erik Panzer for explaining in detail how HyperInt works. We thank Fernando Alday for discussions on large spin bootstrap. We thank IIP-Natal for the hospitality where the final stages of this work was completed.  V.G. would also like to thank FAPESP grant 2015/14796-7, CERN/FIS-NUC/0045/2015 and FCT fellowship SFRH/BD/68313/2010.

\appendix
%%%%%%%%%%%%%%%%%%%%%%%%%%%%%%%%%%%%%%%%%%%%%%%%%%
\section{Conformal integrals}
%%%%%%%%%%%%%%%%%%%%%%%%%%%%%%%%%%%%%%%%%%%%%%%%%%
The skeleton expansion explained in the main text provides an efficient way to express correlation functions of local
operators in terms of Feynman integrals in position space. Ideally one wants to express these integrals in terms of functions that are easy to manipulate, {\em i.e.} take limits of the external parameters, do analytic continuations, evaluate numerically, etc. This is a non-trivial task for generic dimension of space, number of integration variables or powers of the propagators. In the following we shall explain how to obtain the integrals appearing in this work. 

%%%%%%%%%%%%%%%%%%%%%%%%%%%%%%%%%%%%%%%%%%%%%%%%%%%%%%%
\subsection{Series expansion of D-function}
%%%%%%%%%%%%%%%%%%%%%%%%%%%%%%%%%%%%%%%%%%%%%%%%%%%%%%%
The first connected skeleton graph, given by equation (\ref{firstconnectedgraph}),  can be written in terms of a $D$-function. Fortunately, these can be evaluated, in a series expansion of the cross ratios, for generic values of the dimensions of the operators and space 
\be
g(u,v)=C_{\phi\phi\phi}^2\frac{  \Gamma (\Delta )
    }{ 
    \Gamma^4
   \left(\frac{\Delta }{2}\right) \Gamma
   \left(\frac{d}{2}-\Delta \right)}
   u^{\frac{d-\D}{2}}
   \bar{D}_{\frac{d-\D}{2}\,\frac{d-\D}{2}\,\frac{ \D}{2}\,\frac{ \D}{2}}(u,v) \ .
\ee
Using the formulas in \cite{Dolan:2000ut} one can write a double series expansion of the form
\begin{align}
&g(u,v)= \sum_{m,n=0}^\infty u^{m}(1-v)^n
\left[ u^{\frac{\D}{2}}b_{m,n}(d,\D) + u^{\frac{d-\D-2}{2}}c_{m,n}(d,\D)
\right]\\
&= \e \sum_{m,n=0}^\infty u^{m+1}(1-v)^n
\left[  a_{m,n}^{(1,0)}+a_{m,n}^{(1,1)}\ln u+ \e (a_{m,n}^{(2,0)}+a_{m,n}^{(2,1)}\ln u
+a_{m,n}^{(2,2)}\ln^2 u)+O(\e^2)
\right]\nonumber
\end{align}
where the coefficients $a_{m,n}^{(i,j)}$ are known explicitly\footnote{There is a Mathematica file that implements the series expansion of this integral}. 

As we will see, if the powers of the propagators is close to an integer and the dimension $d$ is a even integer then we can use the method that takes advantage of the parametric representation of an integral \cite{Panzer:2013cha,Panzer:2014caa,Panzer:2014gra,Panzer:2015ida}. 

%%%%%%%%%%%%%%%%%%%%%%%%%%%%%%%%%%%%%%%%%%%%%%%%%%%%%%%%%%%%%%%%%
\subsection{Method of differential equations}
%%%%%%%%%%%%%%%%%%%%%%%%%%%%%%%%%%%%%%%%%%%%%%%%%%%%%%%%%%%%%%%%%
At second order in $\epsilon$ one finds the following integral
\begin{align}
\int \frac{d^6x_5d^6x_7}{\pi^6}\frac{1}{
   x_{15}^2 x_{35}^{2}x_{17}^2  x_{37}^{2} (x_{57}^{2}x_{25}^{2}x_{47}^{2})^2  }=\frac{h(z,\bar{z})}{(x_{12}^2x_{34}^2)^2}
\end{align}
Notice that it is simple to obtain a differential equation. It just follows from acting with the Laplacian at position $2$ and using  $\nabla_2^2\frac{1}{(x_2^2)^2}=-4\pi^3\d^{6}(x_2)$
\begin{align}
4u\,\Delta h(z,\bar{z})=4\pi^6 \bar{D}_{1212}(z,\bar{z}),\,\Delta=(1-z)(1-\bar{z})\big[\partial_z\partial_{\bar{z}}+2\frac{\partial_z-\partial_{\bar{z}}}{z-\bar{z}}\big].
\end{align}
The function $D_{1212}$ can be written in terms of a derivative acting on $D_{1111}$ and it turns out that the differential equation for $h(z,\bar{z})$ simplifies if one also writes it in terms of a derivative acting on 
\begin{align}
&4u\Delta \left(\frac{u}{v}\partial_u+\partial_v\right)v \frac{f(z,\bar{z})}{z-\bar{z}} = \left(\frac{u}{v}\partial_u+\partial_v\right)\frac{v u}{z-\bar{z}} \Delta' f(z,\bar{z})=\left(\frac{u}{v}\partial_u+\partial_v\right)v \bar{D}_{1111},\\
&\Delta' = \frac{v}{z-\bar{z}}\partial_{z}\partial_{\bar{z}}.\nonumber
\end{align}
This implies the function $f(z,\bar{z})$ satisfies
\begin{align}
u\Delta'f(z,\bar{z})=\bar{D}_{1111}.
\end{align}
The solution of this equation is given by $\phi^{(2)}(z/(z-1),\bar{z}/(\bar{z}-1))$
\begin{align}
\phi^{(L)}(z,\bar{z})= \sum_{r=0}^{L}\frac{(-1)^r(2L-r)!}{r!(L-r)!L!}\ln^rz\bar{z}\left(\textrm{Li}_{2L-r}\left(z\right)-\textrm{Li}_{2L-r}\left(\bar{z}\right)\right).
\end{align}
Note that the function $D_{1111}$ is nothing more than $\phi^{(1)}(z,\bar{z})$. Thus, we conclude that the function 
\begin{align}
h(z,\bar{z}) = -\left(\frac{u}{v}\partial_u+\partial_v\right)\frac{v}{z-\bar{z}} \phi^{(2)}(z/(z-1),\bar{z}/(\bar{z}-1))
\end{align}

%%%%%%%%%%%%%%%%%%%%%%%%%%%%%%%%%%%%%%%%%%%%%%%%%
\subsection{Integrals in parametric space and how to do them in HyperInt}\label{parametricspaceintegrals}
%%%%%%%%%%%%%%%%%%%%%%%%%%%%%%%%%%%%%%%%%%%%%%%%%
We will focus, for the rest of this section, on a recent method \cite{Panzer:2013cha,Panzer:2014caa,Panzer:2014gra,Panzer:2015ida} that has the advantage of being applicable to 
all the integrals that we have in $d=6+\epsilon$ at least to the order that we have worked out. 

An integral can be represented as a graph $G$ with vertices $V(G)$ that are on the end points of the edges $E(G)$. The vertices can be divided into external points $V_{\textrm{ext}}$ and $V_{\textrm{int}}$. Each edge has associated with it a number $a_e$ which is the power of a specific propagator in the integral
\begin{align}
\Phi(x_i)= \prod_{i\in V_{\textrm{int}}}\bigg(\int\frac{d^dx_i}{\pi^{d/2}}\bigg)\frac{1}{\prod_{i\in V_{\textrm{ext}}\cup V_{\textrm{int}} \,j,\in V_{\textrm{int}}}(x_{ij}^2)^{a_{ij}}}
\end{align}
The parametric representation of an integral uses the concept of tree and forest associated to an integral.  A forest $F \subseteq E(G)$ is defined to be a subgraph without cycles and a tree is a connected forest. The set of connected components of $G$ is denoted as $\pi_0 (G)$. Thus the integral  can be written as \cite{Panzer:2015ida}
\begin{align}
\Phi(x_i) = \frac{\Gamma(\omega)}{\prod_{e\in E}\Gamma(a_{e})} \int \frac{\Omega}{\psi^{d/2}}\left(\frac{\psi}{\phi}\right)^{\omega}\prod_{e\in E} \alpha_{e}^{a_{e}-1},\, \ \ \ \omega=\sum_{e\in E}a_{e}-|V_{\textrm{int}}|d/2\label{eq:SwhingerParametrization}
\end{align}
where $\Omega$ is the measure 
\begin{align}
\int \Omega =\bigg[\prod_{e\in E}\int_{0}^{\infty}d\alpha_{e}\bigg]\delta(1-H(\alpha))
\end{align}
$\psi$ and $\phi$ are defined by
\begin{align}
\psi&= \phi^{P}\ \ \ \ \ \textrm{with} \ \ \ \ P=\{\{v\}:v\in V_{\textrm{ext}}\}\\
\phi &= \sum_{v,w\in V_{\textrm{ext}}} x_{vw}^2\,\Phi^{P_{v,w}} \ \ \ \ \ \textrm{with} \ \ \ \ P_{v,w}=(P \,\textrm{\textbackslash}	 \,\{\{v\},\{w\}\})\cup \{\{v,w\}\}\\
\Phi^{Q}&=\sum_{F}\prod_{e\in F}\alpha_{e}
\end{align}
where the sums are over all spanning forests $F$ of $G$ with $k=|Q|$ connected components $\pi_0(F)=\{T_1,\dots,T_k\}$ such that $P_i\subseteq V(T_i)$ for all $1\le i \le k$. 

The $\psi$ and $\phi$ polynomials for the one loop box integral  (\ref{firstconnectedgraph}) are given by
\begin{align}
\psi=\alpha_1+\alpha_2+\alpha_3\\
\phi=\alpha_2\alpha_3+z\bar{z}\alpha_1\alpha_2+(1-z)(1-\bar{z})\alpha_1\alpha_3 
\end{align}
Notice that the denominator in the integrand parametric representation is linear in the Feynman parameters $\alpha_e$. This makes it possible to write the integral in $\alpha_1$ as a linear combination of hyperlogarithms or Goncharov logarithms with rational prefactors. A given Feynman integral is called linear reducible if there is an order of integration of the Feynman parameters such that at each integration step it can be written in terms of these functions and a rational prefactor whose denominator is linear in the variable being integrated. The package HyperInt \cite{Panzer:2014caa} can be used to check if an integral is linear reducible and it also gives the best integration order\footnote{This can be done using the commands "checkIntegrationOrder" and "suggestIntegrationOrder". } and it also can be used to integrate all Feynman parameters. All integrals appearing in the skeleton expansion of $\phi^3$ at least to oder $\epsilon^3$ are linear reducible. 

The integrals $f$ and $g$ defined in (\ref{eq:3loopintegrals}) can be computed with this method. Their expression is long and consequently we defer its form to an ancilliary file submitted together with this preprint.

The integral defined in (\ref{eq:Twoloopintegralfull}) can be computed explicitly at any order $\epsilon$ with HyperInt. However there is a subtlety since when one of the exponents almost vanishes there is zero coming from the $\Gamma$ functions in the definition of parametric representation. This zero should be canceled by a divergence from the parametric integration. To having to deal with divergences we can just integrate by parts in the corresponding parameter whose propagator goes to zero making the all expression explicitly finite
\begin{align}
\Phi(x_i) = \frac{\Gamma(\omega)}{\prod_{e\in E}\Gamma(a_{e}+\delta_{1e})} \int \prod_{e\in E} \alpha_{e}^{a_{e}-1+\delta_{1e}}\frac{d}{d\alpha_1} \frac{\Omega}{\psi^{d/2}}\left(\frac{\psi}{\phi}\right)^{\omega}
\end{align}
where we assumed that $a_{1}$ goes to zero as $\epsilon\rightarrow 0$. 
%%%%%%%%%%%%%%%%%%%%%%%%%%%%%%%%%%%%%%%%%%%%%%%%%
\section{Conformal blocks}
\label{App:ConfBloEqMo}
%%%%%%%%%%%%%%%%%%%%%%%%%%%%%%%%%%%%%%%%%%%%%%%%%
The conformal blocks resum the contribution of a conformal family to a given four point function. In even space time dimensions they can be written in terms of hypergeometric functions. In this section we study the structure of the conformal blocks in the light-cone limit, {\em i.e.} $u\rightarrow 0$ keeping $v$ fixed,
\begin{align}
G_{\Delta,J}(u,v)=\sum_{m=0}^{\infty} u^{\frac{\Delta-J}{2}+m}g_{m}(v)
\end{align}
with $g_{0}(v)$ given by
\begin{align}
g_0^{\tau,J}(v)  =u^{\frac{\tau}{2}} \left(\frac{v-1}{2}\right)^J\!\!\!\!
 \ _2F_1\! \left(\frac{\tau+2J-\D_{12} }{2} ,\frac{\tau+2J+\D_{34} }{2} , \tau+2J,1-v\right),\,  \ \tau=\Delta-J.
\end{align}
The action of the Casimir operator on this expansion gives a recurrence relation in $m$ for the functions $g_{m}(v)$. The solution for this recurrence relations was found in \cite{Billo:2016cpy}. We show here the value for $m=1$ for the scalar case since these are the ones that are needed to analyze the leading and subleading twist operators
\begin{align}
g_{1}(v)=\frac{\Delta(J+\Delta)^2(\Delta-1)}{(\Delta+J-1)(\Delta+J+1)(\Delta+1-d/2)}\!_2F_1\left(\frac{\Delta+2}{2},\frac{\Delta+2}{2},\Delta+2,1-v\right)
\end{align}
It is possible to see from this last expression the property mentioned in (\ref{eq:FromPrimaryToDescent}). 

%%%%%%%%%%%%%%%%%%%%%%%%%%%%%%%%%%%%%%%%%%%%%%%%%%
\section{$\phi^3$ with global symmetry in adjoint of $SU(N)$}
%%%%%%%%%%%%%%%%%%%%%%%%%%%%%%%%%%%%%%%%%%%%%%%%%%
In this section we will consider $\phi^3$ theory in $d=6+\epsilon$ transforming under the adjoint representation of $SU(N)$. The fields $\phi$ can be represented as, 
\begin{align}
\phi(x)= \phi^{a}(x)T^{a}
\end{align}
where $T^{a}$ is a $N\times N$ matrix transforming in the adjoint of $SU(N)$. The generators of $SU(N)$ satisfy, 
\begin{align}
T^{a}T^{b}=\frac{\d^{ab}\bf{1}_N}{2N}+\frac{1}{2}\sum_{c=1}^{N^2-1}(if^{abc}+d^{abc})T^{c}
\end{align}
where $f^{abc}$ and $d^{abc}$ are antisymmetric and symmetric objects respectively. An interaction term in the Lagrangian like $\Tr(\phi^3)$ will depend on the symmetric object $d^{abc}$. In terms of the generators it can be written as, 
\begin{align}
{\Tr}(T^{a}\big[T^{b},T^{c}\big]_{+})=\frac{d^{abc}}{2}.
\end{align}
where we have used ${\Tr}(T^{a}T^{b})=\delta^{ab}/2$. 
The normalization of this object is, 
\begin{align}
\sum_{c\,,e}^{N^2-1}d_{ace}d_{bce}=\frac{N^2-4}{N}\d_{ab}.
\end{align}
The generators of $SU(N)$ in the adjoint representation satisfy, 
\begin{align}
\sum_{a=1}^{N^2-1}(T^{a})^{i}_{j}(T^{a})^{l}_{k} = \frac{\d_{k}^{i}\d_{j}^l}{2}- \frac{\d_{j}^{i}\d_{k}^l}{2N}.
\end{align}
%\begin{align}
%\sum_{a=1}^{N^2-1}(T^{a})_{ji}(T^{a})_{kl} = \frac{\d_{k}^{i}\d_{j}^l}{2}- \frac{\d_{j}^{i}\d_{k}^l}{2N}.
%\end{align}
The two and three point functions of the field $\phi$ are given by\footnote{A Lagrangian description for this theory is given by, 
\begin{align}
\mathcal{L}=\frac{(\partial_{\mu}\phi^{a})^2}{2}+g\frac{d_{abc}\phi^{a}\phi^{b}\phi^{c}}{3!}
\end{align}}, 
\begin{align}
\left\langle \phi^{a} (x_1)\phi^{b} (x_2)\right\rangle = \frac{\d^{ab}}{(x_{12}^2)^{\D}}\,, \ \ \ \ \ \ 
\left\langle \phi^{a} (x_1)\phi^{b} (x_2)\phi^{c} (x_3)\right\rangle = \frac{C_{\phi\phi\phi}d^{abc}}{(x_{12}^2)^{\frac{\D}{2}}(x_{13}^2)^{\frac{\D}{2}}(x_{23}^2)^{\frac{\D}{2}}}
\end{align}
There are seven representations in the tensor product of two adjoint representations. In the following we will be interested just in the singlet and adjoint channels. According to \cite{Braun:2013tva} the projectors for these two channels are given by
\begin{align}
\mathcal{P}_{\bf{1}}^{a_1a_2a_3a_4}&=\frac{\d^{a_1a_2}\d^{a_3a_4}}{N^2-1}\nonumber\\
\mathcal{P}_{\textrm{adj}}^{a_1a_2a_3a_4}&=\frac{N }{N^2-4}d^{a_1a_2c}d^{a_3a_4c}.
\end{align}
The projectors can be used to decompose the four point function in each irrep channel 
\begin{align}
\mathcal{A}^{a_1a_2a_3a_4}(u,v)=\sum_{R}\mathcal{A}_{R}(u,v)\mathcal{P}^{a_1a_2a_3a_4}_{R}.
\end{align}
Each $\mathcal{A}_{R}(u,v)$ can be obtained using $\sum_{a_i}\mathcal{P}^{a_1a_2a_3a_4}_{R}\mathcal{P}^{a_1a_2a_3a_4}_{R'}=\delta_{R,R'}\textrm{dim}_R$.
In the following we will normalize $\mathcal{A}^{a_1a_2a_3a_4}(u,v)$ such that the identity OPE coefficient of the identity is equal to one.

\bibliography{biblio}

\providecommand{\href}[2]{#2}\begingroup\raggedright\begin{thebibliography}{10}

\bibitem{Fitzpatrick:2012yx}
A.~L. Fitzpatrick, J.~Kaplan, D.~Poland and D.~Simmons-Duffin, \emph{{The
  Analytic Bootstrap and AdS Superhorizon Locality}},
  \href{http://dx.doi.org/10.1007/JHEP12(2013)004}{\emph{JHEP} {\bf 12} (2013)
  004}, [\href{https://arxiv.org/abs/1212.3616}{{\tt 1212.3616}}].

\bibitem{Komargodski:2012ek}
Z.~Komargodski and A.~Zhiboedov, \emph{{Convexity and Liberation at Large
  Spin}}, \href{http://dx.doi.org/10.1007/JHEP11(2013)140}{\emph{JHEP} {\bf 11}
  (2013) 140}, [\href{https://arxiv.org/abs/1212.4103}{{\tt 1212.4103}}].

\bibitem{Alday:2013cwa}
L.~F. Alday and A.~Bissi, \emph{{Higher-spin correlators}},
  \href{http://dx.doi.org/10.1007/JHEP10(2013)202}{\emph{JHEP} {\bf 10} (2013)
  202}, [\href{https://arxiv.org/abs/1305.4604}{{\tt 1305.4604}}].

\bibitem{Cornalba:2007fs}
L.~Cornalba, \emph{{Eikonal methods in AdS/CFT: Regge theory and multi-reggeon
  exchange}},  \href{https://arxiv.org/abs/0710.5480}{{\tt 0710.5480}}.

\bibitem{Caron-Huot:2017vep}
S.~Caron-Huot, \emph{{Analyticity in Spin in Conformal Theories}},
  \href{http://dx.doi.org/10.1007/JHEP09(2017)078}{\emph{JHEP} {\bf 09} (2017)
  078}, [\href{https://arxiv.org/abs/1703.00278}{{\tt 1703.00278}}].

\bibitem{Costa:2012cb}
M.~S. Costa, V.~Goncalves and J.~Penedones, \emph{{Conformal Regge theory}},
  \href{http://dx.doi.org/10.1007/JHEP12(2012)091}{\emph{JHEP} {\bf 12} (2012)
  091}, [\href{https://arxiv.org/abs/1209.4355}{{\tt 1209.4355}}].

\bibitem{Kravchuk:2018htv}
P.~Kravchuk and D.~Simmons-Duffin, \emph{{Light-ray operators in conformal
  field theory}},  \href{https://arxiv.org/abs/1805.00098}{{\tt 1805.00098}}.

\bibitem{Maldacena:2015iua}
J.~Maldacena, D.~Simmons-Duffin and A.~Zhiboedov, \emph{{Looking for a bulk
  point}}, \href{http://dx.doi.org/10.1007/JHEP01(2017)013}{\emph{JHEP} {\bf
  01} (2017) 013}, [\href{https://arxiv.org/abs/1509.03612}{{\tt 1509.03612}}].

\bibitem{Alday:2016njk}
L.~F. Alday, \emph{{Large Spin Perturbation Theory for Conformal Field
  Theories}},
  \href{http://dx.doi.org/10.1103/PhysRevLett.119.111601}{\emph{Phys. Rev.
  Lett.} {\bf 119} (2017) 111601},
  [\href{https://arxiv.org/abs/1611.01500}{{\tt 1611.01500}}].

\bibitem{Gopakumar:2016wkt}
R.~Gopakumar, A.~Kaviraj, K.~Sen and A.~Sinha, \emph{{Conformal Bootstrap in
  Mellin Space}},
  \href{http://dx.doi.org/10.1103/PhysRevLett.118.081601}{\emph{Phys. Rev.
  Lett.} {\bf 118} (2017) 081601},
  [\href{https://arxiv.org/abs/1609.00572}{{\tt 1609.00572}}].

\bibitem{Gopakumar:2016cpb}
R.~Gopakumar, A.~Kaviraj, K.~Sen and A.~Sinha, \emph{{A Mellin space approach
  to the conformal bootstrap}},
  \href{http://dx.doi.org/10.1007/JHEP05(2017)027}{\emph{JHEP} {\bf 05} (2017)
  027}, [\href{https://arxiv.org/abs/1611.08407}{{\tt 1611.08407}}].

\bibitem{Dey:2016mcs}
P.~Dey, A.~Kaviraj and A.~Sinha, \emph{{Mellin space bootstrap for global
  symmetry}}, \href{http://dx.doi.org/10.1007/JHEP07(2017)019}{\emph{JHEP} {\bf
  07} (2017) 019}, [\href{https://arxiv.org/abs/1612.05032}{{\tt 1612.05032}}].

\bibitem{Dey:2017oim}
P.~Dey and A.~Kaviraj, \emph{{Towards a Bootstrap approach to higher orders of
  epsilon expansion}},
  \href{http://dx.doi.org/10.1007/JHEP02(2018)153}{\emph{JHEP} {\bf 02} (2018)
  153}, [\href{https://arxiv.org/abs/1711.01173}{{\tt 1711.01173}}].

\bibitem{Dey:2017fab}
P.~Dey, K.~Ghosh and A.~Sinha, \emph{{Simplifying large spin bootstrap in
  Mellin space}}, \href{http://dx.doi.org/10.1007/JHEP01(2018)152}{\emph{JHEP}
  {\bf 01} (2018) 152}, [\href{https://arxiv.org/abs/1709.06110}{{\tt
  1709.06110}}].

\bibitem{Henriksson:2018myn}
J.~Henriksson and M.~Van~Loon, \emph{{Critical O(N) model to order $\epsilon^4$
  from analytic bootstrap}},  \href{https://arxiv.org/abs/1801.03512}{{\tt
  1801.03512}}.

\bibitem{Alday:2015ewa}
L.~F. Alday and A.~Zhiboedov, \emph{{An Algebraic Approach to the Analytic
  Bootstrap}}, \href{http://dx.doi.org/10.1007/JHEP04(2017)157}{\emph{JHEP}
  {\bf 04} (2017) 157}, [\href{https://arxiv.org/abs/1510.08091}{{\tt
  1510.08091}}].

\bibitem{Mack:1973kaa}
G.~Mack, \emph{{Conformal invariance and short distance behavior in quantum
  field theory}}, \href{http://dx.doi.org/10.1007/BFb0017087}{\emph{Lect. Notes
  Phys.} {\bf 17} (1973) 300--334}.

\bibitem{Petkou:1994ad}
A.~Petkou, \emph{{Conserved currents, consistency relations and operator
  product expansions in the conformally invariant O(N) vector model}},
  \href{http://dx.doi.org/10.1006/aphy.1996.0068}{\emph{Annals Phys.} {\bf 249}
  (1996) 180--221}, [\href{https://arxiv.org/abs/hep-th/9410093}{{\tt
  hep-th/9410093}}].

\bibitem{Petkou:1995vu}
A.~C. Petkou, \emph{{C(T) and C(J) up to next-to-leading order in 1/N in the
  conformally invariant 0(N) vector model for 2 < d < 4}},
  \href{http://dx.doi.org/10.1016/0370-2693(95)00936-F}{\emph{Phys. Lett.} {\bf
  B359} (1995) 101--107}, [\href{https://arxiv.org/abs/hep-th/9506116}{{\tt
  hep-th/9506116}}].

\bibitem{Petkou:1996np}
A.~C. Petkou, \emph{{Operator product expansions and consistency relations in a
  O(N) invariant fermionic CFT for 2 < d < 4}},
  \href{http://dx.doi.org/10.1016/S0370-2693(96)01227-0}{\emph{Phys. Lett.}
  {\bf B389} (1996) 18--28}, [\href{https://arxiv.org/abs/hep-th/9602054}{{\tt
  hep-th/9602054}}].

\bibitem{Dolan:2000ut}
F.~A. Dolan and H.~Osborn, \emph{{Conformal four point functions and the
  operator product expansion}},
  \href{http://dx.doi.org/10.1016/S0550-3213(01)00013-X}{\emph{Nucl. Phys.}
  {\bf B599} (2001) 459--496},
  [\href{https://arxiv.org/abs/hep-th/0011040}{{\tt hep-th/0011040}}].

\bibitem{Panzer:2015ida}
E.~Panzer, \emph{{Feynman integrals and hyperlogarithms}}.
\newblock PhD thesis, Humboldt U., Berlin, Inst. Math., 2015.
\newblock \href{https://arxiv.org/abs/1506.07243}{{\tt 1506.07243}}.

\bibitem{Anselmi:1998ms}
D.~Anselmi, \emph{{The N=4 quantum conformal algebra}},
  \href{http://dx.doi.org/10.1016/S0550-3213(98)00848-7}{\emph{Nucl. Phys.}
  {\bf B541} (1999) 369--385},
  [\href{https://arxiv.org/abs/hep-th/9809192}{{\tt hep-th/9809192}}].

\bibitem{Henn:2005mw}
J.~Henn, C.~Jarczak and E.~Sokatchev, \emph{{On twist-two operators in N=4
  SYM}}, \href{http://dx.doi.org/10.1016/j.nuclphysb.2005.09.043}{\emph{Nucl.
  Phys.} {\bf B730} (2005) 191--209},
  [\href{https://arxiv.org/abs/hep-th/0507241}{{\tt hep-th/0507241}}].

\bibitem{Rychkov:2015naa}
S.~Rychkov and Z.~M. Tan, \emph{{The $\epsilon$-expansion from conformal field
  theory}}, \href{http://dx.doi.org/10.1088/1751-8113/48/29/29FT01}{\emph{J.
  Phys.} {\bf A48} (2015) 29FT01},
  [\href{https://arxiv.org/abs/1505.00963}{{\tt 1505.00963}}].

\bibitem{deAlcantaraBonfim:1980pe}
O.~de~Alcantara~Bonfim, J.~Kirkham and A.~McKane, \emph{{Critical Exponents to
  Order $\epsilon^3$ for $\phi^3$ Models of Critical Phenomena in Six
  $\epsilon$-dimensions}},
  \href{http://dx.doi.org/10.1088/0305-4470/13/7/006}{\emph{J.Phys.} {\bf A13}
  (1980) L247}.

\bibitem{Gracey:2015tta}
J.~A. Gracey, \emph{{Four loop renormalization of $\phi^3$ theory in six
  dimensions}}, \href{http://dx.doi.org/10.1103/PhysRevD.92.025012}{\emph{Phys.
  Rev.} {\bf D92} (2015) 025012}, [\href{https://arxiv.org/abs/1506.03357}{{\tt
  1506.03357}}].

\bibitem{Alday:2017vkk}
L.~F. Alday and S.~Caron-Huot, \emph{{Gravitational S-matrix from CFT
  dispersion relations}},  \href{https://arxiv.org/abs/1711.02031}{{\tt
  1711.02031}}.

\bibitem{Alday:2016jfr}
L.~F. Alday, \emph{{Solving CFTs with Weakly Broken Higher Spin Symmetry}},
  \href{http://dx.doi.org/10.1007/JHEP10(2017)161}{\emph{JHEP} {\bf 10} (2017)
  161}, [\href{https://arxiv.org/abs/1612.00696}{{\tt 1612.00696}}].

\bibitem{Panzer:2013cha}
E.~Panzer, \emph{{On the analytic computation of massless propagators in
  dimensional regularization}},
  \href{http://dx.doi.org/10.1016/j.nuclphysb.2013.05.025}{\emph{Nucl. Phys.}
  {\bf B874} (2013) 567--593}, [\href{https://arxiv.org/abs/1305.2161}{{\tt
  1305.2161}}].

\bibitem{Panzer:2014caa}
E.~Panzer, \emph{{Algorithms for the symbolic integration of hyperlogarithms
  with applications to Feynman integrals}},
  \href{http://dx.doi.org/10.1016/j.cpc.2014.10.019}{\emph{Comput. Phys.
  Commun.} {\bf 188} (2015) 148--166},
  [\href{https://arxiv.org/abs/1403.3385}{{\tt 1403.3385}}].

\bibitem{Panzer:2014gra}
E.~Panzer, \emph{{On hyperlogarithms and Feynman integrals with divergences and
  many scales}}, \href{http://dx.doi.org/10.1007/JHEP03(2014)071}{\emph{JHEP}
  {\bf 03} (2014) 071}, [\href{https://arxiv.org/abs/1401.4361}{{\tt
  1401.4361}}].

\bibitem{Billo:2016cpy}
M.~Billo, V.~Goncalves, E.~Lauria and M.~Meineri, \emph{{Defects in conformal
  field theory}}, \href{http://dx.doi.org/10.1007/JHEP04(2016)091}{\emph{JHEP}
  {\bf 04} (2016) 091}, [\href{https://arxiv.org/abs/1601.02883}{{\tt
  1601.02883}}].

\bibitem{Braun:2013tva}
V.~M. Braun and A.~N. Manashov, \emph{{Evolution equations beyond one loop from
  conformal symmetry}},
  \href{http://dx.doi.org/10.1140/epjc/s10052-013-2544-1}{\emph{Eur. Phys. J.}
  {\bf C73} (2013) 2544}, [\href{https://arxiv.org/abs/1306.5644}{{\tt
  1306.5644}}].

\end{thebibliography}\endgroup
\bibliographystyle{JHEP}

\end{document}